\documentclass[epj,referee]{svjour}
\usepackage{verbatim,float}
\usepackage{graphics}
\usepackage{graphicx}
\usepackage{subfigure}
\usepackage{dcolumn}
\usepackage{bm}
\usepackage{color} 
\usepackage{hyperref}
\usepackage{epstopdf}
\usepackage{booktabs}
\usepackage{balance}

\begin{document}

\title{Second-Order Phase Transition Induced by Deterministic
	Fluctuations in  Aperiodic Eight-State Potts 
	Models}
	
\author{Pradheesh R\inst{1} \and Harikrishnan S. Nair\inst{2} \and 
	V. Sankaranarayanan\inst{1} \and K. Sethupathi\thanks{Author for correspondence 
	(rpradeesh@physics.iitm.ac.in).}}
	 
\institute{Low Temperature Physics Laboratory, Department of Physics, Indian Institute of Technology Madras, Chennai 600036, India.. 
 	\and 
 	J\"{u}lich Center for Neutron Sciences 2/ Peter Gr\"{u}nberg Institute 4, 
Forschungszentrum J\"{u}lich GmbH, 52425 J\"{u}lich, Germany. 
	}

\date{22 June 1998}
%
%
%

\makeatother

\newenvironment{decl}
  {\par\small\addvspace{1.9ex plus 1ex}%
    \vskip -\parskip
    \noindent\hspace{-\leftmargini}%
    \begin{tabular}{|l|}\hline\ignorespaces}
  {\\\hline\end{tabular}\nobreak%
    \vspace{2.2ex}\vskip -\parskip}
\title{Large Magnetoresistance and Jahn-Teller Effect in Sr$_2$FeCoO$_6$}

\author{Pradheesh R. \inst{1} $^{\ast}$,\and Harikrishnan S. Nair \inst{2},\and V. Sankaranarayanan \inst{1} $^{\ast}$ \and K. Sethupathi\inst{1} $^{\ast}$}

\institute{$^1$Low Temperature Physics Laboratory, Department of Physics, Indian Institute of Technology Madras, Chennai 600036, India. \\
$^2$ J\"{u}lich Center for Neutron Sciences 2/ Peter Gr\"{u}nberg Institute 4, 
Forschungszentrum J\"{u}lich GmbH, 52425 J\"{u}lich, Germany. \\
}
\date{\today}
%
\abstract{
Neutron diffraction measurement on the spin glass double perovskite Sr$_2$FeCoO$_6$ reveals site disorder as well as Co$^{3+}$ intermediate spin state.
 In addition, multiple valence states of Fe and Co are confirmed through M\"{o}ssbauer and X-ray photoelectron spectroscopy.
The structural disorder and multiple valence lead to competing ferromagnetic and antiferromagnetic interactions and subsequently to a spin glass state, which is reflected in the form of an additional $T$-linear contribution at low temperatures in specific heat.
A clear evidence of Jahn-Teller distortion at the Co$^{3+}$-O$_6$ complex is observed and incorporating the physics of Jahn-Teller effect, the presence of localized magnetic moment is shown.
A large, negative and anomalous magnetoresistance of $\approx$ 63$\%$ at 14~K in 12~T applied field
is observed for Sr$_2$FeCoO$_6$.
The observed magnetoresistance could be explained by applying a semi-empirical fit consisting of a negative and a positive contribution and show that the negative magnetoresistance is due to spin scattering of carriers by localized magnetic moments in the spin glass phase.
}
\maketitle
%
\section{Introduction}
%
The rich physical properties of $A_2MM^{\prime}$O$_6$ ($A$ = alkaline earth;
$M,M'$ = transition metal) double perovskites
are well documented in the review article by Serrate {\it et. al.,}
\cite{serrate2007double}.
One of the most-studied system in this respect is the ferromagnet Sr$_2$FeMoO$_6$ (SFMO)
\cite{sarma_2000}, 
which exhibits magnetoresistance at elevated temperature of about  420~K.
Structurally ordered as well as disordered SFMO
were studied by Sarma {\it et al.,}
\cite{sarma_2000}
who observed the existence of sharp changes in low-field
MR in the case of ordered sample.
Cobalt-doped Sr$_2$FeMoO$_6$ also exhibits an appreciable MR (40 $\%$) which was attributed to the inhomogeneous magnetic state
\cite{chang_2006}.
Structural defects like antisite disorder, antiphase boundaries
and oxygen stoichiometry lead to competing magnetic phases
in double perovskites
\cite{sarma_2000,ogaleapl_75_537_1999}.
For example, magnetic spin glass phase stemming from intrinsic disorder
was reported in nanoscale powders of Sr$_2$FeMoO$_6$
\cite{poddar_jap_106_073908_2009evidence}.\\
Most of the experimental investigations on double perovskite oxides where focused on compounds in which $M$ and $M^{\prime}$ were a combination of first and third row transition metals (TM).
Recently, the studies aimed at combining the first-row TMs (Co and Mn, for example) have pointed towards the importance of valence states of the TMs in these systems
\cite{baidya_2011}.
For example, Mn$^{3+}$ and Co$^{3+}$ can have intermediate spin states (IS) which can lead to Jahn-Teller (JT) distortions where the JT ion will have a single $e_g$ electron and the double degeneracy of $e_g$ state will be lifted by the distortion
\cite{kugel_1982}.
This will lead to itinerant behaviour of the single electron which will contribute to magnetic and transport properties.
\\
There are only a few reports on experimental investigations on the Co based double perovskite, Sr$_2$FeCoO$_6$
\cite{maignan_3_57_2001,bezdicka_91_501_1994}.
{\it Ab~initio} band structure calculations on Sr$_2$FeCoO$_{6}$
showed that both Co and Fe make comparable contributions to ferromagnetism
of the compound
\cite{bannikov2008}
where Co$^{4+}$ and Fe$^{4+}$ in high spin states
can lead to metallicity and ferromagnetism.
However, owing to the comparable ionic radii and
valence states of Fe and Co (both in 4+ state), it is surprising
that this material showed ferromagnetism.
Recently, Co-based double perovskites have attracted
interest also due to the excellent electrochemical performance
as anode materials in solid oxide fuel cells
\cite{wei_jmc_22_225_2011cobalt}.
In our previous investigation on the magnetic properties of Sr$_2$FeCoO$_6$,
\cite{pradheesh}
clear signatures of spin glass state originating from structural disorder
and multiple valence states of Fe/Co were obtained which motivated us to
investigate its possible consequences on the magnetoresistive properties.
In the present article we report a comprehensive study on the
structure, M\"{o}ssbauer and X-ray photoelectron spectroscopy, specific heat
and magnetoresistance of Sr$_2$FeCoO$_6$.%
\section{Experimental}
%
%
Sr$_2$FeCoO$_6$ samples used in this study were
prepared by sol-gel method as described elsewhere
\cite{pradheesh,deac_2002}.
Neutron powder diffraction (NPD) measurements at 12, 30, 60, 65 and 300~K were performed
with neutron wavelength of 1.4789 \mbox{\AA} employing position sensitive detector.
The crystal structure was refined by Rietveld method
\cite{rietveld}
using FULLPROF program
\cite{carvajal}.
M\"{o}ssbauer measurements in transmission mode were performed
with $^{57}$Co source in constant acceleration mode using a M\"{o}ssbauer spectrometer equipped with a Weissel velocity drive.
The spectrum was collected at room temperature using Canberra S-100 constant-acceleration M\"{o}ssbauer spectrometer having the $^{57}$Co in Rh matrix.
The velocity was calibrated using the spectra of $\alpha$-Fe foil.
The software Fit;O was used to analyze the M\"{o}ssbauer spectrum
\cite{fito_2006}.
Electron core level binding energy was analyzed
using Philips X-ray photoelectron spectrometer (PHI5700).
%
%
%
\begin{figure}[!t]
\label{structure}
\begin{center}
\includegraphics[width=0.3\textwidth]{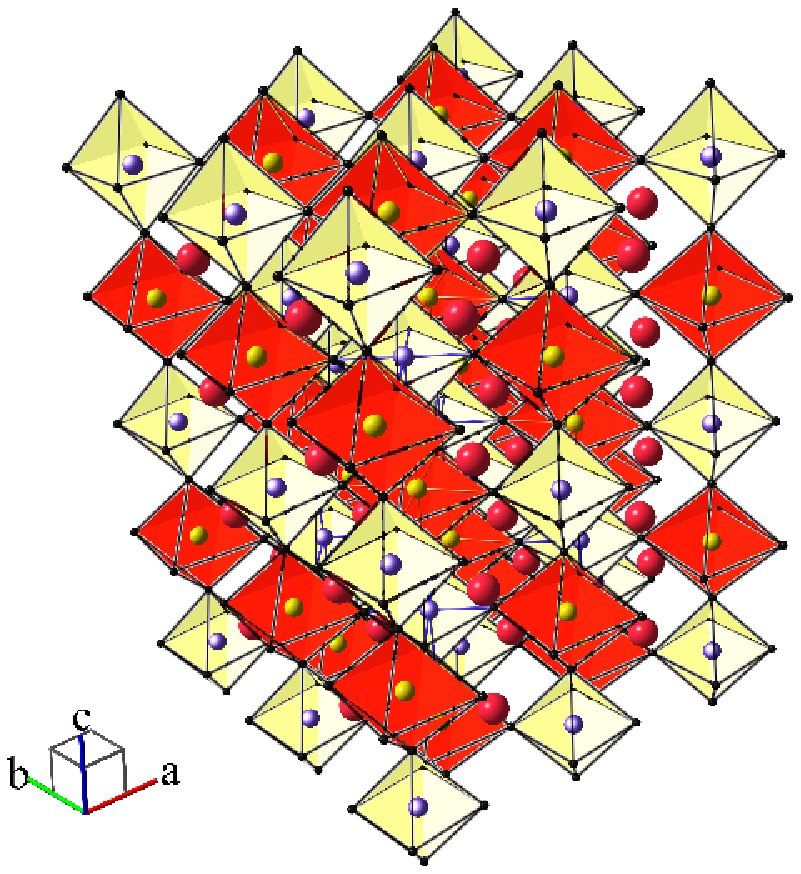}\\
\includegraphics[width=0.25\textwidth]{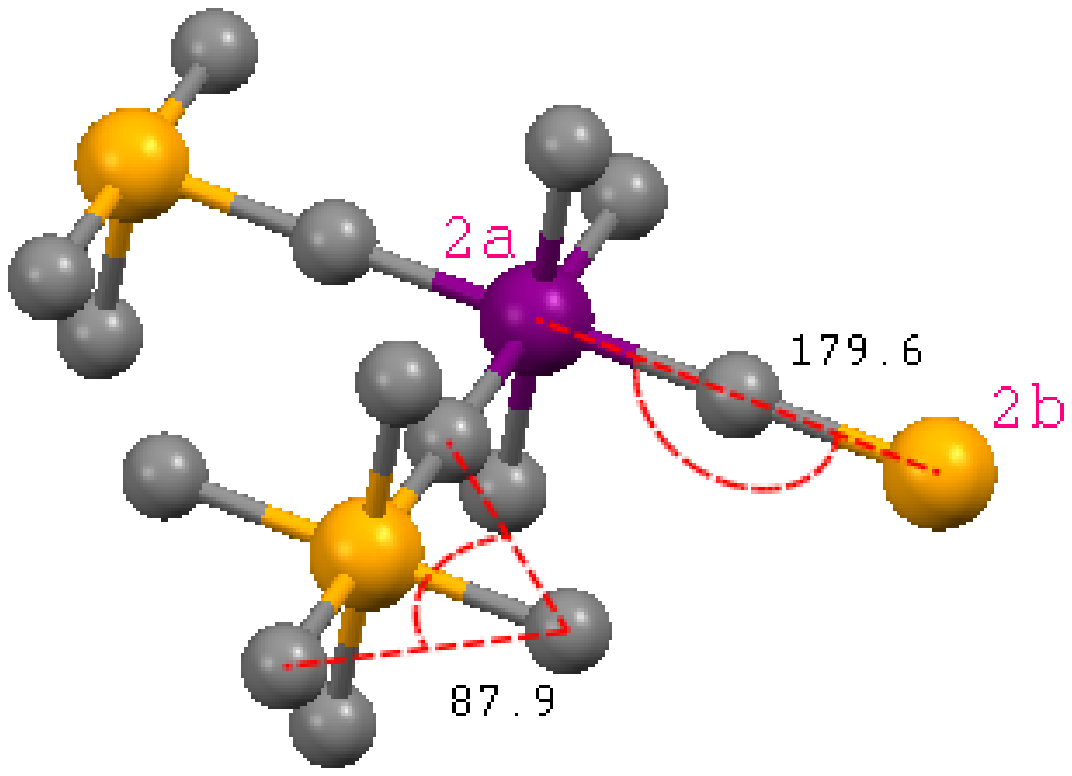}
\end{center}
\caption{(colour online) Top: A schematic of the crystal structure  of Sr$_2$FeCoO$_6$. Bottom: FeO$_6$ and CoO$_6$ octahedra are shown, where Fe are orange and Co purple. The tilt ($\phi_1$=179.64) and rotation ($\phi_2 = 87.94$) angles at 12~K are also shown.}
\end{figure}
Specific heat measurements were performed using a semi-adiabatic heat pulse
technique in the temperature range of 4--300~K at 0~T and applied field of 8~T.
Electrical resistivity in 0~T and applied magnetic field of 6~T was measured employing standard four-probe method on circular pellets of Sr$_2$FeCoO$_6$.
Magnetoresistance measurements were carried out using a 12~T superconducting magnet
in an exchange cryostat.
\section{Results and Discussion}
\subsection{Neutron diffraction}
\begin{figure}[!h]
\centering
\includegraphics[scale=0.95]{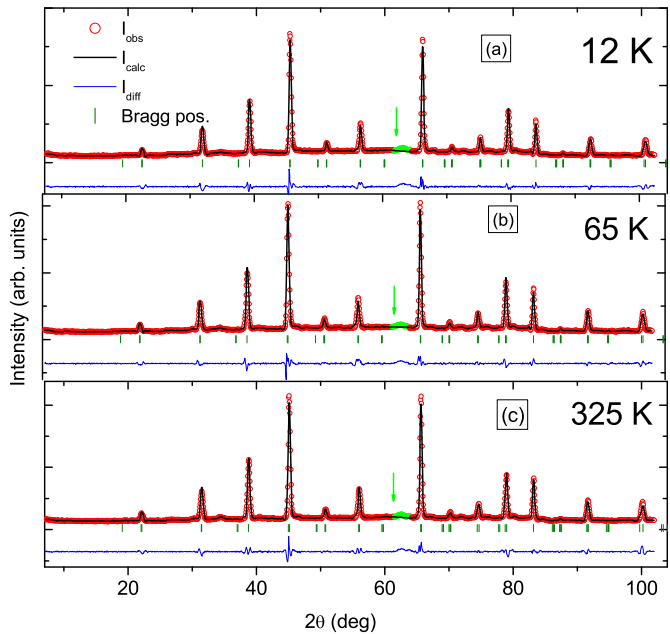}
\caption{(colour online) Neutron powder diffraction patterns of Sr$_2$FeCoO$_6$ at (a) 12~K, (b) 65~K and (c) 325~K are presented along with the results of Rietveld refinement. The structure was refined in $I4/m$ space group at all temperatures. In the figure, a vertical down-arrow (coloured green) indicates the region of 2$\theta$ = 61.4 to 64.1$^{\circ}$  which was excluded from the refinement.}
\label{ND}
\end{figure}

The analysis of neutron diffraction data was performed using tetragonal $I4/m$ space group, presented in figure~\ref{ND}.
%
\begin{table}[!ht]
\caption{Structural parameters of Sr$_2$FeCoO$_6$ at different temperatures. The Wyckoff positions are Sr $4d$(0,$\frac{1}{2}$,$\frac{1}{4}$); Fe1 $2a$(0,0,0); Fe2 $2b$(0,0,$\frac{1}{2}$); Co1 $2b$(0,0,$\frac{1}{2}$); Co2 $2a$(0,0,0); O$_{ap}$ $4e$(0,0,$z$) and O$_{eq}$ $8h$ ($x$, $y$, $z$). The agreement factors of the refinement are quantified as $\chi^2$ and $R_{Bragg}$.}
\centering
\scalebox{0.8}{
\begin{tabular}{llllll} \hline\hline\
            & & 12~K   & 30~K & 65~K &  300~K \\ \hline\hline
Space Group && $I4/m$ & $I4/m$ & $I4/m$ & $I4/m$    \\
$a (\mbox{\AA})$   && 5.4360(1) & 5.4390(2)& 5.4381(1)& 5.4554(3)    \\
$c (\mbox{\AA})$   && 7.7010(3) & 7.6982(3)& 7.7037(3)& 7.7319(1)    \\
$V (\mbox{\AA}^3)$ && 227.569(3)& 227.736(2)& 227.826(1)& 230.115(5)  \\
O$_{ap}$        & $4e$ (0, 0, z) &     &     &     &           \\
$z$         && 0.2592(1)& 0.2502(2)& 0.2495(1)& 0.2530(4) \\
O$_{eq}$        & $8h$ (x, y, 0) &     &     &      &            \\
$x$         && 0.2491(9)& 0.2548(1)& 0.2581(9)& 0.2498(2) \\
$y$         & &0.2507(6)& 0.2433(7)& 0.2400(3)& 0.2541(7) \\
$\chi^2$    &  & 3.5 & 3.4 & 3.37 & 3.83            \\
$R_{Bragg}$   & & 4.2 & 4.2 & 3.97 &  3.63       \\ \hline\hline
\end{tabular}}
\label{tab1}
\end{table}
%
\balance
A completely random occupancy of Fe and Co in the $2a$ (0,0,0)
and $2b$ (0,0,$\frac{1}{2}$) sites, as previously done in the case of X-ray investigations
\cite{pradheesh},
were assumed in the refinement procedure.
The refined lattice parameters, unit cell volume and atomic positions
of Sr$_2$FeCoO$_6$ (SFCO) determined from the Rietveld analysis are presented in Table~\ref{tab1}.
A completely disordered arrangement of cations
were also reported in the case of double perovskites like Sr$_2$MnNbO$_6$, Sr$_2$MnRuO$_6$
\cite{lufaso_jssc_177_1651_2004},
where $I4/mcm$ space group also was suggested for
a completely disordered cationic arrangement.
Hence, refinement of the structure in $I4/mcm$ space group
with Sr($4b$) (0,$\frac{1}{2}$,$\frac{1}{4}$); Fe/Co($4c$)
(0,0,0); O($4a$) (0,0,$\frac{1}{4}$); O($8h$) ($x$,$x+\frac{1}{2}$,0)
was also performed for SFCO.
%
%
When the octahedral volume is occupied
by 50:50 mixture of cations, the most probable tilt system
is $a^0 a^0 a^-$ and the space group being $I4/m$ if the 
cations are ordered or $I4/mcm$ if they are disordered.
%
However, in our analysis reproduction of the observed intensities
and better agreement factors were accomplished with
the choice of $I4/m$.\\
Previous macroscopic magnetic measurements on Sr$_2$FeCoO$_6$
\cite{pradheesh}
revealed that the valence state of Co was +3 and the spin
state as IS (intermediate spin).
IS Co$^{3+}$ is a Jahn-Teller active ion and 
hence it distorts the Co--O$_6$ octahedra to further remove the degeneracy of the triply degenerate t$_{2g}$ and doubly degenerate e$_g$ states.
The experimentally determined bond lengths for SFCO, given in Table~\ref{tab2},
 show the presence of two long and four short bond lengths for CoO$_6$
and two short and four long bonds for FeO$_6$ 
suggesting the distortion of CoO$_6$
octahedra owing to Jahn-Teller (JT) activity.
%
%
\begin{table}[!ht]
\caption{Main bond distances (\mbox{\AA}) and selected angles (deg) of Sr$_2$FeCoO$_6$ at different temperatures. The distortion parameter of the octahedra is defined as $\Delta_d$ = (1/6)$\Sigma_{n=1-6}[(d_n - <d>)/<d>]^2$. $<d>$ is the average Mn--O distance. The strain parameter is $e_t = (c_p - a_p)/(c_p + a_p)$. $c_p$ and $a_p$ are the equivalent primitive perovskite lattice constants given by $c$/2 and $a$/$\sqrt(2)$ respectively. The structural parameters
in the absence of distortions as calculated using SPuDS are also shown.}
\centering
\begin{tabular}{lllllll} \hline\hline
       & 12~K & 30~K & 65~K & 300~K &  \\ \hline\hline
Co--O$_{ap}$ (2)& 1.9961(2)& 1.9267(1)  & 1.9221(6)& 1.9227(5)&            \\
Co--O$_{eq}$ (4) & 1.9217(3)& 1.9164(5) & 1.913(4) & 1.9107(5)&            \\
Fe--O$_{ap}$ (2) & 1.8543(5)& 1.9224(1) & 1.9297(1)& 1.9329(5)&             \\
Fe--O$_{eq}$ (4) & 1.9221(5)& 1.9305(2) & 1.934(4)&  1.9695(5)&             \\
$\langle$Fe--O--Co$\rangle$ & 179.64(1)& 177.38(1)& 176.04(2)& 177.92(1)&            \\
$\Delta_d \times$10$^{-4}$ & 3.25 & 0.064 & 0.77 & 0.52& \\ 
$c_p/a_p$ & 1.001 & 1.0 & 1.001 & 1.002 &  \\ 
$e_t \times$10$^{-4}$ & 8.7 & 4.1 & 8.4 & 13& \\
$\phi_1$(deg) & 0.18 & 1.31 & 1.98 & 1.04 & \\
$\phi_2$(deg) & 1.03 & 0.84  & 0.07 & 1.94&  \\ 
\hline
\end{tabular}
\begin{tabular}{llllll}
SPuDS  &  $a (\mbox{\AA})$    &   $c (\mbox{\AA})$    & Fe--O    & Co--O  &  $\langle $Fe--O--Co$\rangle$ \\ \hline
           & 5.3684      & 7.5921       & 1.9260     & 1.8700   & 179.99   \\ \hline\hline
\end{tabular}
\label{tab2}
\end{table}
%
%
\balance
The bond distances Fe--O increase with increasing temperature
while Co--O distances show the opposite trend.
The Jahn-Teller distortion has three sets of modes namely, breathing mode ($Q_1$), basal distortion ($Q_2$) and octahedral stretching ($Q_3$).
From the bond length calculation, two long and four short, it is obvious that Co$^{3+}$ shows $Q_3$ distortion mode \cite{satpathy_1996}.
The distortions of the octahedra are reflected in the octahedral
tilting of $M$O$_6$ from the $z$ axis defined as $\phi_1$ = (180 - $\Theta$)/2
where $\Theta$ is the $M$--O--$M'$ bond angle along $z$ axis.
The rotation angle in the $xy$-plane, defined
as $\phi_2$ = (90 - $\Omega$)/2 where $\Omega$ is the angle among three oxygen ions between two corner-shared oxygen octahedra, also quantifies the degree of distortion.
The tilt angle $\phi_1$ and the rotation angle $\phi_2$ of SFCO at different temperatures are given in Table~\ref{tab2}.
The values of average bond angles $<$Fe-O-Co$>$ and the distances of Fe--O and Co--O bonds and various distortion parameters are also presented in the table.
They are not as large compared to the values observed in double perovskites like Sr$_2$MnSbO$_6$
\cite{cheah_jssc_179_1775_2006jahn}
however, $\phi_1$ ($\phi_2$) shows high (low) values in the intermediate temperature range
where an anomaly in lattice constants were reported earlier
\cite{pradheesh}.
The octahedral tilting and the distortions are also quantified through
empirical parameters like, $c_p/a_p$ ($c_p$ = $c$/2 and $a_p$ = $a$/$\sqrt 2$) which denote the ratio of equivalent primitive perovskite lattice parameters; and the distortion parameter $\Delta_d$, defined as (1/6)$\Sigma_{n=1-6}[(d_n - <d>)/<d>]^2$.
Typical values of these parameters for SFCO are given in Table~\ref{tab2}.
From Table~\ref{tab2}, it is clear that the tetragonal strain defined as, $e_t = (c_p - a_p)/(c_p + a_p)$, also shows an increase with increasing temperature.
In order to signify the distortions of the octahedra,
we have calculated the lattice parameters
and the bond parameters at room temperature using the program SPuDS
\cite{lufaso_acta_57_725_2001prediction}
(see, Table~\ref{tab2}).
A comparison between the ideal values calculated by SPuDS 
and the experimentally determined values for SFCO
indicates the major role played by lattice distortions in this
double perovskite.
%
\subsection{M\"{o}ssbauer and X-ray Photoelectron Spectroscopy}
%
%
\begin{figure}[!h]
\centering
\includegraphics[scale=0.45]{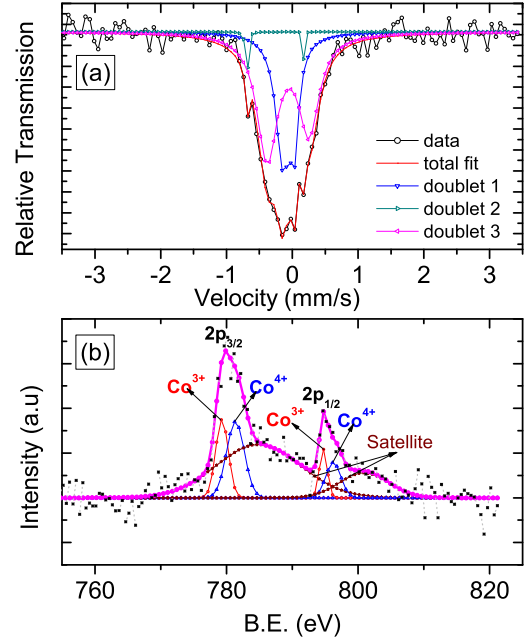}
\caption{(color online) (a) Room temperature M\"{o}ssbauer spectra of Sr$_2$FeCoO$_6$. The experimental data are shown in black circles and the total fit assuming three doublets is presented as red solid line. Separate contributions from the three doublets are also shown. (b) The Co-XPS spectra of Sr$_{2}$FeCoO$_{6}$ and the fits for Co$^{3+}$ and Co$^{4+}$.
The black dashed line and squares represents experimental data points and the
solid magenta line represents the fitted curve.}
\label{Moss_xps}
\end{figure}
%
Cationic disorder and mixed valence of the transition
metals Fe and Co results in inhomogeneous
magnetic exchange interactions leading to magnetic
frustration and spin glass state in SFCO.
In our earlier work, using structural and bulk magnetic investigations, has been confirmed the presence of spin glass state in this material
\cite{pradheesh}.
However, a clear picture regarding the valence and spin states of cations has not emerged.
A clear understanding regarding the valency of the cations and
ligand coordination can be obtained from M\"{o}ssbauer
experiments since it probes the local environment of the cations
and has been successfully used as a tool to explore
the magnetism and electronic structure of the colossal magnetoresistive perovskites
\cite{nemeth_jnrs_9_R1_2008}.
The experimentally obtained room temperature M\"{o}ssbauer
spectrum of SFCO is presented in Fig.~\ref{Moss_xps}(a).
The spectra shows paramagnetic behaviour and is deconvoluted
in to three doublets with different isomer shifts and quadrupole
splitting.
On the basis of theoretical calculations using the difference in the electron density
at the absorber and source, it was determined that the Fe$^{4+}$ in low-spin (LS) state will
have an isomer shift ($\delta$) value in the range 0.1--0.2 mm/s
\cite{dickson_1986}.
The isomer shift, $\delta$, and quadrupole splitting of the paramagnetic
components, $\Delta$, determined using the least square fit are given
in Table~\ref{arttype}, from which we found that the isomer
shift for doublet 2 is in good agreement
with that of low-spin Fe$^{4+}$ and doublet 3
for Fe$^{3+}$ in an octahedral coordination
\cite{shimony_1966}.
In comparison, the $\delta$ value of doublet 1 is intermediate to
that of Fe$^{4+}$ and Fe$^{3+}$ ($>$0.30 mm/s) and hence
is attributed to Fe$^{3+y}$ due to the rapid ion exchange as commonly
observed in ferrite structures at ordinary temperatures
\cite{motohashi_2006}.
Thus, possibility of a charge disproportionation of the form
2Fe$^{4+} \rightarrow$ Fe$^{(4-d)+}$ + Fe$^{(4+d)+}$
cannot be ruled out
\cite{homonnay_2002}.
For a comparison, M\"{o}ssbauer spectra of Sr$_2$FeTeO$_6$ \cite{martinjmc_16_66_2006} was fitted to
two symmetric doublets with equal isomer shift
values of 0.435(2) and 0.434(1)~mm/s and quadrupolar splitting of
0.23 and 1.21~mm~s$^{-1}$ which are consistent with high spin Fe$^{3+}$.
%
%
\begin{table}
\caption{\label{arttype} The M\"{o}ssbauer parameters, isomer shift ($\delta$), quadrupolar shift ($\Delta$) and the area ratio of Sr$_2$FeCoO$_6$ at room temperature.}
\centering
\begin{tabular}{c c c c} \hline
Doublet              &  $\delta$ (mm s$^{-1}$)      &   $\Delta$ (mm s$^{-1}$)     &   Area Ratio\\ \hline\hline
1  & 0.1788(3)   &  0.2009(2)  &50.58$\%$    \\
2  & 0.1118(2)   &  0.6322(2)  &28.26$\%$     \\
3  & 0.2809(3)   &  0.5524(1)  &21.16$\%$      \\ \hline\hline
\end{tabular}
\end{table}
%
The scenario of multiple valence states for Fe and Co in SFCO is also
supported by X-ray photoelectron spectroscopy (XPS).
Figure~\ref{Moss_xps} (b) presents the experimentally observed XPS spectrum of  Co in SFCO deconvoluted into
Gaussian components which results in an apparent doublet
for Co $2p_{1/2}$ and Co $2p_{3/2}$.
From the fit, the doublet positions are obtained at $779.64(3)$
and $781.36(3)$~eV for Co $2p_{1/2}$ and $794.86(4)$ and $796.24(3)$~eV for
Co $2p_{3/2}$ with shake-up peaks at $788.63$~eV and $800.70$~eV.
The binding energies of $779.64(3)$~eV  and $794.86(4)$~eV are attributed
to Co$^{3+}$ while $781.36(3)$~eV and $796.24(3)$~eV to Co$^{4+}$ since
Co$^{4+}$ has greater electron density than Co$^{3+}$
\cite{liu_2009}.
The presence of shake-up satellite for Co$^{3+}$ confirms that it is not
in a low spin state since LS has $t_{2g}$ states completely
filled minimizing the possibility of metal to ligand charge transfer
and shake-up peaks.
\begin{table}
\caption{\label{xps}Binding energy (BE) of Co 2$p$ peaks in Sr$_2$FeCoO$_6$ compared with other cobaltites. Here, Oct. stands for octahedral environment while LBE and HBE are lower binding energy (2$p_{3/2}$) and higher binding energy (2$p_{1/2}$). Co is in octahedral coordination in all compounds.}
\centering
\scalebox{0.8}{
\begin{tabular}{c c c c c c c}
\hline
&\multicolumn{3}{c}{B.E.}\\
\cline{3-4}
Oxide              & 	 Ion  &		$2p_{3/2}$    &  $2p_{1/2}$  & Reference\\ \hline
Sr$_2$FeCoO$_6$  &  	3+  & 779.61 (3)  &  794.86(4)  & present work    \\
LiCoO$_2$  &  	3+  &	779.5  & 794.6  & \cite{oku_1978}    \\
Co$_3$O$_4$  &  	3+  &	779.6  & 794.5  & \cite{chuang_1976}    \\
La$_{1-x}$Sr$_x$CoO$_{3-\delta}$  &  3+  &	779.6  & 794.9  & \cite{tabata_1987}    \\
LaCoO$_3$  & 	3+  &	780   & 794.5  & \cite{munakata_1997} \\
Sr$_2$FeCoO$_6$  &  	4+  & 781.36 (3)	  &  796.24 (3)  & present work \\
Ba$_{0.5}$Sr$_{0.5}$Co$_{0.8}$Fe$_{0.2}$O$_{3-\delta}$  &  	4+  &	780.29  & 795.63  & \cite{liu_2009}    \\ \hline\hline
\end{tabular}}
\end{table}
%
\balance
Table~\ref{xps} shows a comparison of the valence states
estimated for SFCO with those of other perovskite cobaltites.
Based on the results from M\"{o}ssbauer and XPS experiments and combining
our previous results from magnetic measurements, we can
arrive at the possible combinations of different valence states for Fe and Co in SFCO.
%
\begin{figure}[!b]
\centering
\includegraphics[trim = 6 2 2 2, clip=true, scale=0.75]{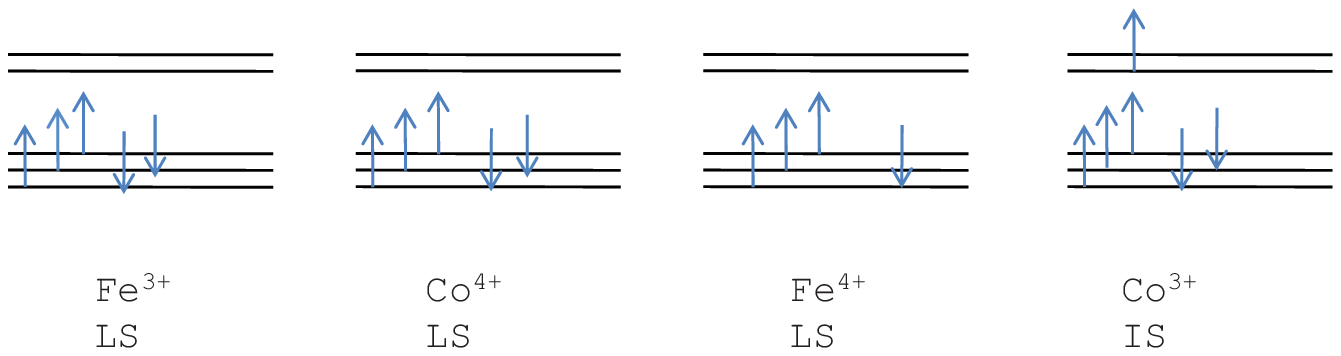}
\caption{(color online) Schematic of the level schemes of various spin state for different valences of Fe and Co in Sr$_2$FeCoO$_6$. The spin states are indicated
as $IS$ = intermediate spin, $LS$ = low spin.}
\label{SS}
\end{figure}
In Fig.~\ref{SS} (a), a schematic of the possible spin states that
explains the properties of SFCO is presented.
\subsection{Specific Heat}
\begin{figure}
\centering
\includegraphics[scale=0.45]{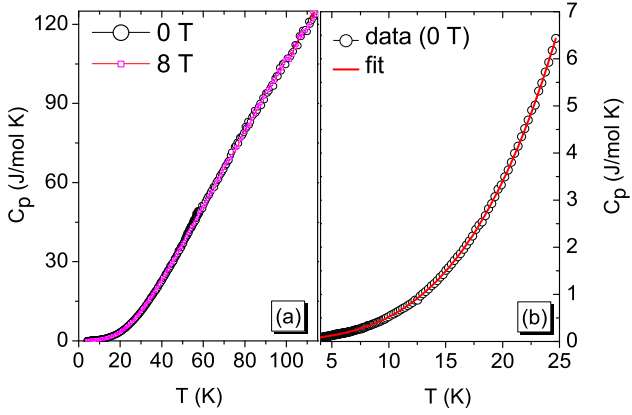}
\caption{(a) The specific heat $C_p(T)$ of Sr$_{2}$FeCoO$_{6}$ in zero magnetic field and an applied field of 8~T. (b) The low temperature part of the specific heat at 0~T along with the fit using eqn.~\ref{fit1}.}
\label{spheat}
\end{figure}

%
The temperature dependent specific heat, $C_p(T)$, in the range 4 -- 120~K at an applied field of 0~T as well as 8~T is presented in Fig.~\ref{spheat} (a) .
A high value of specific heat, comparable in magnitude to that observed in insulating
manganites Dy$_{0.5}$Sr$_{0.5}$MnO$_3$
\cite{hari_jpcm_20_275234_2008},La$_{0.8}$Ca$_{0.2}$MnO$_3$
\cite{hamilton_prb_54_14964_1996},
is observed in SFCO also.
%
No anomaly corresponding to the spin glass transition at
$T_c \sim$ 75~K \cite{pradheesh}
is observed in $C_p(T)$ which clearly marks the absence
of any collective thermodynamic transition or long range order.
It is clear from the figure that even with the application of
8~T there is no appreciable change in the specific heat.
We have tried to model the experimentally observed specific heat,
and a satisfactory fit to the experimental data
was obtained using a model that includes a linear term in $T$
which accounts for the spin glass contribution, as represented by eqn.~\ref{fit1}.
The fit achieved in the temperature range 4--25~K is shown in Fig.~\ref{spheat} (b).
\begin{equation}
C (T) = \gamma T+\beta_3T^3+\beta_5 T^5
\label{fit1}
\end{equation}
Spin glass systems show a strong linear dependence of specific heat at low temperatures
\cite{mydosh_1993,martin_1979}
and moreover the linear dependence was verified using a full
mean field solution of a quantum Heisenberg spin glass model in large $N$ limit
\cite{georgesprl_2000}.
The disordered nature and the magnetic glassy state, as in the case of manganites
\cite{banerjee_2009},
also contribute to linear $T$ dependence of specific heat.
%
\begin{table}
\caption{\label{CP} The fit-parameters obtained from the specific heat analysis performed using eqn.~\ref{fit1}.}
\centering
\begin{tabular}{|l||l|} \hline\hline
$\gamma$ (J mol$^{-1} $K$^{-2}$)       &    0.01655(2)                       \\
$\beta_3$ (J mol$^{-1}$ K$^{-4}$)    &    3.4916(1)$\times$10$^{-4}$  \\
$\beta_5$ (J mol$^{-1}$ K$^{-6}$)    &    8.5383(1)$\times$10$^{-8}$   \\
$\Theta_D$ (K)                                   &   409 \\ \hline\hline
\end{tabular}
\end{table}
Resistivity measurements, discussed below (Fig.~\ref{MR} (a)), show that SFCO  is an insulator
with the low temperature resistance
reaching a value of the order of $10^6$ $\Omega$cm and hence the $T$--linear term that
we observe in $C_p(T)$ originates from the spin degrees of freedom and disorder.
The coefficient of linear term, $\gamma$, obtained from the analysis was
16 mJ mol$^{-1}$ K$^{-2}$ which is close to 11 mJ mol$^{-1}$ K$^{-2}$ observed in the case of
La-doped Sr$_2$FeMoO$_6$
\cite{tovar_jmmm_272_857_2004}.
The Debye temperature $\Theta_D$ = 409~K was estimated for SFCO using the relation $\Theta_D$=$(12\pi^4pR/5\beta_3)^{1/3}$, where $R$ is
the gas constant and $p$ is the number of atoms.
The specific heat coefficient $\beta_3$ is extracted from the fit.
The Debye temperature estimated is comparable to the value $\Theta_D$ = 463~K
\cite{pradheesh} from Gr\"{u}neisen approximation for the unit cell volume.
The specific heat investigations on Sr$_2$FeMoO$_6$, for example, have reported a $\Theta_D$ value of 338~K
\cite{okuda_prb_14_144407_2003}
which lies close to the value obtained for SFCO.
A detailed analysis of the specific heat data of Sr$_2$FeTeO$_6$ using
two different Debye temperatures for the light and heavy atoms resulted
in the values of 768~K and 261~K respectively
\cite{martinjmc_16_66_2006}.
Attempts to analyze the $C_p(T)$ data including a $T^{3/2}$ term in order to
take in to account the antiferromagnetic spin wave
\cite{esrgopal_1966}
were also made.
However, the fitting was unfavourable due to the
presence of powers of $T$ with comparable magnitudes.
The value of $\gamma$ obtained
from the fit is due to the spin glass nature and is
comparable to the values observed in spin glass
\cite{wiebeprb_68_13_2003}
and cluster glass perovskites
\cite{gusmaossc_127_9_683}.
%
\subsection{Magnetoresistance}
\begin{figure}
\centering
\includegraphics[scale=0.35]{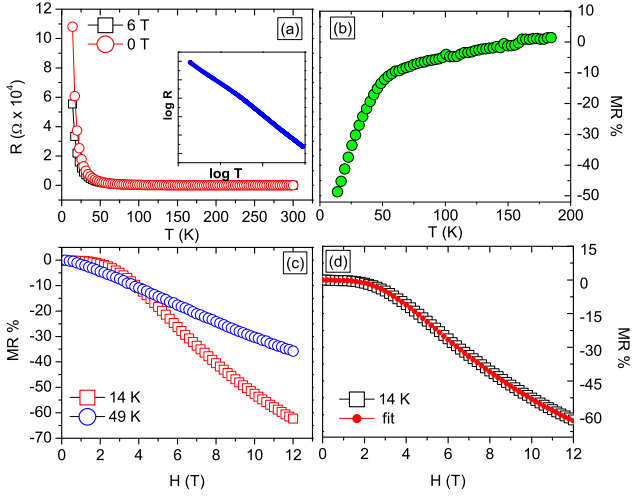}
\caption{(colour online) (a) Temperature dependence of electrical resistance of Sr$_{2}$FeCoO$_{6}$ in
zero field and an applied field of 6~T. Inset shows the zero field resistance in log-log scale.
(b) Magnetoresistance of Sr$_2$FeCoO$_6$ as a function of temperature. (c) Magnetoresistance  as a function of applied magnetic field at two temperatures, 14~K and 49~K.
(d) Fit to the isothermal MR at 14~K using eqn.~(\ref{fit2}). The fit parameters obtained are,
$A_1$= 0.1086(4), $A_2$= 0.0013(4), $B_1$ = 0.0159(3) and $n$ =1.43(6).}
\label{MR}
\end{figure}

%
The electrical resistance of SFCO in 0 and 6~T applied magnetic
field is presented in Fig.~\ref{MR} (a). It exhibits a monotonic increase as
the temperature is lowered below the spin glass transition temperature.
The variation of electrical resistance indicates semiconducting behaviour.
The inset of the figure shows the log-log representation of the main
panel and suggests that thermally activated behaviour of carrier
hopping does not hold in SFCO.
The resistivity does not follow Arrhenius law and a real energy gap opening being the reason for the upturn of resistance at low temperature is ruled out since no divergence in the derivative of resistivity is observed. 
%
Colossal magnetoresistance effects are normally observed at
the transition temperatures (metal-insulator or magnetic)
but in the case of SFCO, maximum MR is found at very low temperatures.
The temperature evolution of MR is presented in Fig.~\ref{MR} (b).
The MR is positive for a small range at high temperatures and then begins
to decrease with decreasing temperature and goes to negative values.
Negative magnetoresistance observed in semiconductors was
compared with the effects observed in dilute alloys such as
Cu-Mn and a theory based on localized magnetic moments
was proposed by Toyazawa
\cite{toyozawajpsj_1962}.
Later, in an empirical model suggested by Khosla and Fischer
\cite{khoslaprb_1970},
a negative component to magnetoresistance stemming
from the localized magnetic moment model of Toyazawa and a positive component
due to the field dependence of Hall coefficient were combined
to explain the data on CdS.
In order to explain the observed magnetoresistance in SFCO,
we employ the empirical relation derived by Khosla and Fischer.
Thus, the isothermal MR was found to fit to the following expression,
\begin{equation}
\frac{\Delta\rho}{\rho}= - {A_1}^2\ln(1 + {A_2}^2H^2)+\frac{{B_1}^2H^n}{1+{B_2}^2H^2}
\label{fit2}
\end{equation}
where $A_1, A_2, B_1, B_2$ and $n$ are free parameters of fitting and H is the applied magnetic field.
$A_1$ and $A_2$ are given by
\begin{eqnarray}
{A_1}^2= {a}^2J \left[S(S+1) + \left\langle M^2 \right\rangle\right] \\
{A_2}^2 = \left[1 + 4{S}^2{\pi}^2{\left(\frac{2J \rho_F}{g}\right)}^4\right]\frac{g^2\mu^2}{\left({\alpha}k_B T\right)^2}
\label{par1}
\end{eqnarray}
where  $a$ is regarded as a measure of spin scattering.
The positive component of MR is proportional to $H^n$ where $n$ ranges from  $0.77$ to $2.19$.
A value less than $1.8$ has been observed for systems showing a monotonic increase in
negative MR
\cite{luoepjb_2006}.
The field-dependent MR of SFCO at 14~K and 49~K are presented
in Fig.~\ref{MR} (c) and the data at 14~K along with the fit using
equation~(\ref{fit2}) are presented in Fig.~\ref{MR} (d).
The value of $n$ = 1.43(6) obtained from this fit lies in the range consistent with negative MR.
\\
Owing to crystal field effects, the five-fold degenerate 3$d$ level of Co$^{3+}$ splits into three-fold degenerate $t_{2g}$ and two-fold degenerate $e_g$ states.
The system may prefer further removal of degeneracy through Jahn-Teller distortion, if it can help in lowering the energy.
As schematically shown in Fig ~\ref{SS}, only Co$^{3+}$ has an occupied $e_g$ state.
It suits the system to undergo Jahn-Teller distortion of the Co$^{3+}$--O$_6$ octahedra.
Our neutron diffraction measurement confirms the Co$^{3+}$--O$_6$ distortion as shown in Fig ~\ref{structure}.
The distortion is manifested by changing the tilt angle and rotation angle from $180^\circ$ and $90^\circ$ respectively.
However, the presence of a Jahn-Teller ion can lead to a local structural distortion thereby  removing the degeneracy of the two-fold $e_g$ state.
If, through such a degeneracy removal, the upper level of the two-fold $e_g$ states is available to an electron, it can strongly influence the conduction mechanism of SFCO.
From the structural studies it is evident that there is an increase in the bond length along the apical direction which means a descent in the symmetry to $D_{4h}$.
The lowering of the symmetry is treated as a perturbation to the cubic field and in order to link the local structural distortion to the energy level splitting, we note that the energy separation between the  $d_{x^2 - y^2}$ and $d_{z^2}$ arises from the second order crystal field coefficient of distorted octahedra.
Hence the Jahn-Teller coupling $E_{JT}$ can be described as
\cite{ballhausen_1962}
\begin{eqnarray}
E_{JT}^{e_g} = 4D_s + 5D_t \\
E_{JT}^{t_{2g}} = 3D_s - 4D_t \\
D_s  = \frac{e}{14} \sqrt{\frac{5}{\pi}}A_{20}\left\langle r^{2}\right\rangle_{3d}
\label{D1} \\
D_t  = \frac{e}{14\sqrt{\pi}}T\left\langle r^{4}\right\rangle_{3d} 
\label{D2}
\end{eqnarray}
where $A_{20}$ is the second order crystal electric field (CEF) coefficient, 
$T = A^{\prime}_{40} - A_{40}$ is the difference between fourth order CEF coefficients of distorted octahedra and perfect octahedra.
$\left\langle r^n \right\rangle$ is the expectation value of the $n^{th}$ power of the radial distance of a $3d$ orbital from the nucleus.
$D_s$ and $D_t$ are splitting parameters for the spherical harmonics $Y^0_2$ and $Y^0_4$ as given in eqn ~(\ref{Ds}) and eqn ~(\ref{Dt}) and $R_{3d}$ is the normalized spherically symmetric radial functions, $r$ is the radius and $R$ is the Co--O bond length.
\begin{eqnarray}
   D_s = \int[R_{3d}(r)]^2\frac{3}{2}\frac{r^2}{R^3}.r^2dr
   \label{Ds}\\
   D_t = \int[R_{3d}(r)]^2\frac{3}{2}\frac{r^4}{R^5}.r^2dr
   \label{Dt}
\end{eqnarray}
Using eqn.~(\ref{Ds}) and eqn.~(\ref{Dt}) we can arrive at eqn.~(\ref{D1}) and eqn.~(\ref{D2}).
Taking the radius of IS Co$^{3+}$ (0.56 \mbox{\AA}), the value of $D_t$ works out to be less than $D_s$.
Hence the main contribution to energy arises from $A_{20}$ and $E_{JT}$ can be written as
\begin{equation}
  E_{JT}^{e_g} \approx 4D_s =\frac{2e}{7}\sqrt{\frac{5}{\pi}}A_{20}\left\langle r^2 \right\rangle_{3d}
\end{equation}
The radius of Co$^{3+}$ in IS state is nearly equal to that of Fe$^{3+}$ (0.55 \mbox{\AA}) and hence the second order CEF coefficient can be evaluated using the quadrupole splitting obtained from the  M\"{o}ssbauer data.
The quadrupole splitting can be expressed as
$\Delta = \frac{eQV_{zz}}{2}\left(1+\frac{\eta^2}{3}\right)^{1/2}$
where $Q$ is the electric quadrupole moment of the $^{57}$Fe nucleus, $\eta$ is the asymmetry parameter, and $V_{zz}$ is the principal component of the electric field gradient at the nucleus which contains contributions from valence as well as the surrounding electrons.
\begin{eqnarray}
V_{zz}= V_{zz}(Fe) + V_{zz}(lattice) \\
V_{zz}(lattice) = \sum \frac{q_i\left(3cos^2\theta - 1 \right)}{R_i^3} = \sqrt{\frac{5}{\pi}}A_{20}
\end{eqnarray}
Combining eqns (11) and (13) and converting $Q$ to eV by a factor of $\frac{E\gamma}{c}$, where $c$ is the velocity of light and using $Q$ = 0.28$\times$10$^{-24}$ cm$^2$, $\eta$ = 0 for axial symmetry we get $E_{JT}^{e_g}$ = 1.3~eV (note that this is the minimum energy and hence a contribution from $D_t$ will only increase the energy).
This value is high enough such that the electron in the $e_g$ level is not itinerant but localized, consistent with the experimentally observed resistivity.
The large energy separation E$_{JT}$, traps the electron leading to localization;
consequently, the ground state is insulating.
%
\begin{figure}
\includegraphics[scale=0.40]{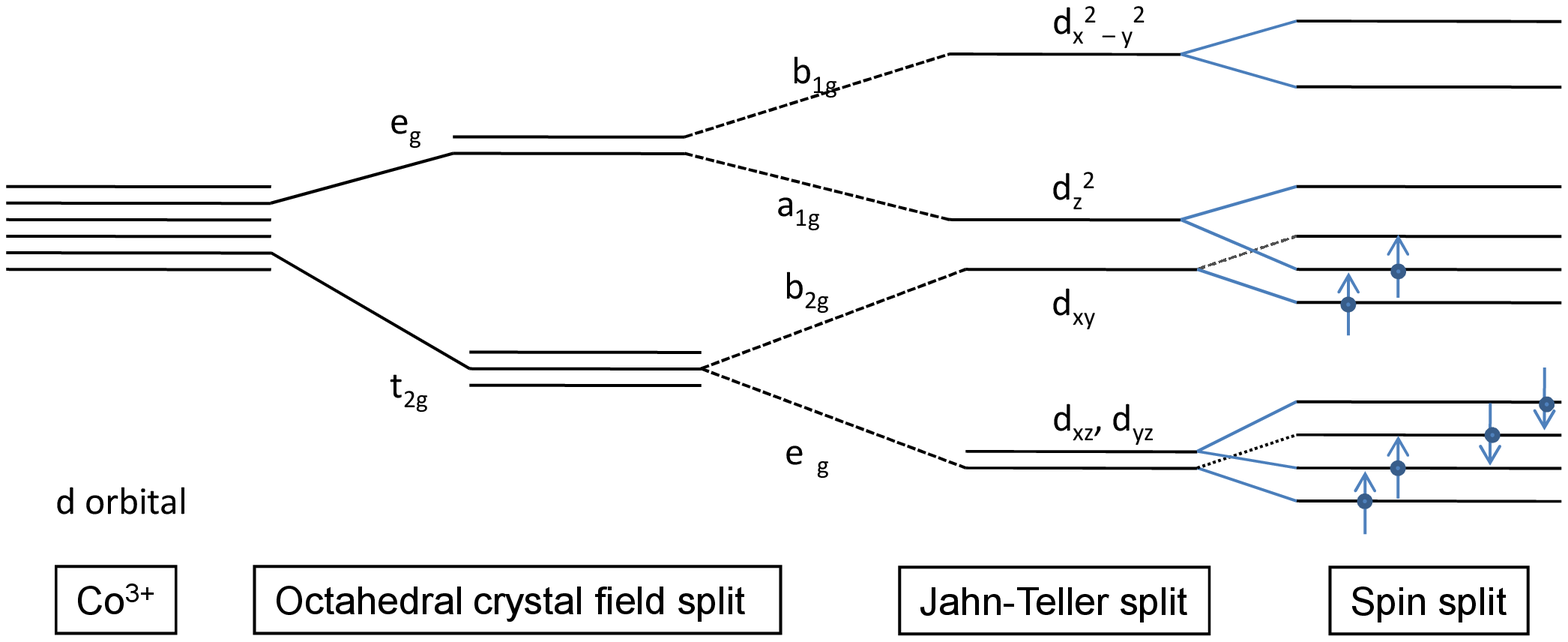}
\caption{(colour online) Schematic of Co$^{3+}$ level scheme
 in IS state showing the removal of degeneracy.}
\label{IS}
\end{figure}%
%
From Fig ~\ref{SS} we can see that all the other valence states have unoccupied $e_g$ levels and as the $t_{2g}$ states are already localized due to crystal field effect the Co--O--Fe interaction is antiferromagnetic.
Due to the tetragonal symmetry, the three-fold $t_{2g}$ state will further split
into an $b_{2g}$ state and a doubly degenerate $e_g^\prime$ state while the $e_g$
state will split into $a_{1g}$ and $b_{1g}$ states as shown in Fig.~\ref{IS}.
The insulating behaviour of SFCO suggests that the conducting carriers
are holes that lie mainly in $a_{1g}$ state
\cite{mizokawa_2001}.
The holes are strongly coupled to lattice and leads to localization resulting in localized magnetic moments due to spin-phonon coupling.
\\
A small contribution of positive MR is also seen in the present case which 
is associated with the spin glass behaviour observed in this system.
Strong competition occurs between spin clusters and ferromagnetic state below the spin glass transition temperature.
As an external magnetic field is applied the ferromagnetic coupling strength get enhanced and hence we observe
a large magnetoresistance.
The mixed valence  and disordered arrangement of Fe and Co leads to complex magnetic
behaviour forming the spin glass state.
From neutron diffraction and magnetization measurements it was confirmed that Fe and Co in SFCO have mixed valence states.
While Co$^{3+}$ is in an intermediate state the other valence states are in low spin states.
Preferential stability of the IS state over HS due to strong ligand metal hybridization has also been
reported already
\cite{saitoh_1997},\cite{louca_1999}.
Hence we infer that the positive contribution to MR, though small, is due to the coexistence of the FM and AFM interactions.
As seen in SFCO the magnetoresistive behaviour due to the enhancement of the FM regions due to the application of a magnetic field in a cluster glass system were also reported
in the Co-rich phases of Co doped Sr$_2$FeMoO$_6$
\cite{chang_2006}.
It is instructive to compare the negative magnetoresistive value of 63$\%$ in SFCO with the case of
Sr$_2$FeMoO$_6$ which shows 42$\%$ MR and Sr$_2$CoMoO$_6$
with zero MR
\cite{viola_cm_14_812_2002}.
%
\section{Conclusions}
%
To summarize, the neutron diffraction study and subsequent analysis of the structural parameters confirm the tetragonal $I4/m$ structure and reveal Jahn-Teller distortions present in SFCO double perovskite.
Based on the analysis of M\"{o}ssbauer and XPS spectra we substantiate the presence of mixed valency of transition metal cations which leads to an inhomogeneous magnetic state in SFCO. 
These studies also confirm the intermediate spin state of Co$^{3+}$ in this material.
Large magnetoresistance of 63$\%$ is observed at low temperatures in SFCO.
Using Khosla-Fischer model, it is confirmed that the MR has negative and positive components
which originates from the localized magnetic moments and the spin glass state respectively.
%
\section*{Acknowledgements}
The authors are grateful to B. R. K. Nanda for his comments
 and critical review in improving the manuscript.
The magnetoresistance measurements were performed at IGCAR Kalpakkam.
Specific heat measurements were performed at UGC-DAE Consortium for Scientific Research, Indore, India. 
The authors thank A. Bharathi and A. T. Satya, IGCAR Kalpakkam for their help in MR measurements. 
PR wishes to thank Rajeev Rawat, Ajay Gupta, R. Nirmala, K. Balamurugan  and E. Senthil Kumar for stimulating discussions. \\
\newpage

%

%
%
%
%
%
%
%
%
\end{document}